# Multi-scale dynamical symmetries and selection rules in nonlinear optics

Gavriel Lerner[1,2], Ofer Neufeld[1,2], Liran Hareli[3], Georgiy Shoulga[3], Eliayu Bordo[1,2], Avner Fleischer[4], Daniel Podolsky[1], Alon Bahabad[3] and Oren Cohen[1,2]

[1]Physics Department, Technion – Israel Institute of Technology, Haifa, Israel.
[2]Solid State Institute, Technion – Israel Institute of Technology, Haifa, Israel.
[3]Department of Physical Electronics, Tel Aviv University, Tel Aviv, Israel.
[4]Chemistry Department, Tel Aviv University, Tel Aviv, Israel.
Corresponding authors' e-mail addresses: gavriel@campus.technion.ac.il, oren@si.technion.ac.il

**Symmetries and their associated selection rules are extremely useful in all fields of science. In particular, for systems that include electromagnetic (EM) fields interacting with matter, it has been shown that both the symmetries of matter [1,2] and the EM field's time-dependent polarization [3,4] play a crucial role in determining the properties of the linear and nonlinear responses. The relationship between the system's symmetry and the properties of its excitations facilitate precise control over light emission [5,6] and enable ultrafast symmetry-breaking spectroscopy of a variety of properties [7,8]. Here, we formulate the first general theory that describes the macroscopic and microscopic dynamical symmetries (including quasicrystal-like symmetries) of an EM vector field, revealing many new symmetries and selection rules in light-matter interactions. We demonstrate an example of multi-scale selection rules experimentally in the framework of high harmonic generation (HHG). This work paves the way for novel spectroscopic techniques in multi-scale systems as well as for imprinting complex structures in EUV-X-ray beams, attosecond pulses, or the interacting medium itself.**

Symmetry is regularly used to derive conservation laws and to formulate forbidden/allowed transitions in interacting systems [1]. In the field of nonlinear optics, symmetries are standardly used to determine whether a particular nonlinear process is allowed or forbidden according to the medium's point-group [2,9]. Recently, a more general group theory was developed describing the symmetries of the EM field's time-dependent polarization (denoted dynamical symmetries [3], DSs) and its interaction with matter [4]. Such DSs and their associated selection rules have been applied to shaping the waveforms of EUV and X-ray radiation emitted from HHG [10–12] and enabled ultrafast symmetry-breaking spectroscopy of molecular [13,14] and solid orientation [15], molecular symmetries [14], and chirality [7,8]. However, this theory of DSs is local (operating solely on a microscopic scale) [4], and thus fully neglects light's macroscopic structure. Moreover, it does not account for composite microscopic-macroscopic (multi-scale) DSs.

Here, we formulate a general theory for EM fields and their interactions with matter where the multi-scale symmetries of the full light-matter Hamiltonian are analyzed. We describe spatio-temporal DSs as generalized unitary transformations, and all possible symmetry operations that close under group multiplication are systematically derived. Various combinations of EM fields are cataloged into different groups that are comprised of one or more DSs. We assign each DS an associated selection rule that indicates the allowed frequencies, polarizations, momenta, and angular momentum of the harmonic emission. Our theory generalizes many previous results, such as complex structured XUV emission generated by spatio-temporal structure beams in the beams longitudinal axis [5,16–21] or profile [6,10–12]. We also discovered new types of symmetries, including simultaneous spin-orbit angular momentum conservation and the periodic [22] and aperiodic [23] space-time crystals of a vector field. We explore several new multi-scale DSs numerically and experimentally in the framework of HHG to demonstrate the richness of this approach for light-matter interactions.

We begin by describing the multi-scale DSs of an EM vector field, which are combinations of temporal, microscale and macroscale spatial building blocks. Then, we derive a general equation that determines the selection rules of the polarization and frequencies (temporal and spatial) of high harmonics.

**Multi-scale symmetries**

We analyze here the symmetries of an arbitrary EM vector field; however, the theory is general and can be applied to analyzing the symmetries of time-periodic Hamiltonians and other equations of motion, as we show in later sections. The electric and magnetic components of EM waves in isotropic media exhibit the same DS; hence, we consider here only the electric field. The basic entity we explore is the vector field:

$$\vec{E} = \vec{E}(\vec{R}, t) = \vec{E}(\vec{X}) \tag{1}$$

where $\vec{R}$ denotes the spatial-dependence of this field and $t$ represents its temporal-dependence. For brevity, we define a general spatio-temporal coordinate vector, $\vec{X}$. It is beneficial to separate the three types of DOF of $\vec{E}(\vec{X})$: (i) As a vector field, $\vec{E}(\vec{X})$ has three independent polarization components – $E_x$, $E_y$ and $E_z$. We denote these as **microscopic DOF** of the field, as they reflect its intrinsic local structure in a given spatial location. (ii) $\vec{E}$ may depend on three spatial coordinates. We denote these dependencies as **macroscopic DOF**, as they reflect the spatial structure of the field. (iii) $\vec{E}$ depends on the **time** coordinate. A symmetry of $\vec{E}$ is an operation that keeps it invariant; hence, a complex spatio-temporal operation $\hat{G}$ is a symmetry if $\hat{G} \cdot \vec{E} = \vec{E}$. The 'order' of this operation is the number of times it needs to be applied to itself until it returns to the identity, i.e., it is $n$th order if $n$ is the first integer for which $\hat{G}^n = 1$.

The field $\vec{E}$ can exhibit a large variety of symmetries depending on the role played by its microscopic and macroscopic DOF. First, the polarization components could have intricate dependencies amongst themselves along the time-axis, resulting in microscopic symmetries. These DSs have recently been fully described by a local Floquet group theory [4], where general DSs involve products of temporal operations (either time translations by time $T/n$ where $T$ is the temporal period, denoted by $\hat{t}_n$, or time-reversal denoted by $\hat{T}$, see table 1 of the SI) and local operations (summarized in table 2 of the SI) that intermix the polarization components and are equivalent to point-group symmetries (local rotations by $2\pi m/n$ around axis $j$, denoted by $\hat{r}^j_{n,m}$; local reflections, denoted by $\hat{\sigma}$; local inversion, denoted by $\hat{\iota}$; and local improper rotations by $2\pi m/n$, denoted by $\hat{s}_{n,m}$). Second, the field could have symmetries associated with its macroscopic structure and spatial-dependence as well as its temporal behavior [22]. In this case, the spatial symmetry operations comprise the space-group symmetries which include the latter operators denoted by capital letters (these are not to be confused with the lower-case letter operators, which act on the local field polarization components). Besides point-group operations, spatial translations are also possible (where translation by $Lm/n$ along the $j$ axis is denoted by $\hat{J}_{n,m}$, and $L$ is the minimal spatial period along the $j$'th axis). Lastly, a periodic space-time (but aperiodic in time or space) field is invariant under a mixed operation of macrospace-time translations, $\widehat{D}_n$. We emphasize that the operations acting on the microscopic and macroscopic scales need not be the same ones. Altogether, these options give rise to a rich and diverse theory that describes the symmetries of $\vec{E}$, with ($N$+$M$+$1$) DOF, where $N$ is the number of local DOF, $M$ is the number of macroscopic DOF, and "$1$" is the time dimension. The recent local Floquet group theory spans the (3+0+1) case [4,24], while the group theory for time-dependent crystals is described by the (0+3+1) case [22]. For ordered non-periodic systems, e.g., quasicrystals [25] and optical quasicrystals [26,27], we use super-space concepts [23] where $M$ is larger than the physical macroscopic dimension.

We can systematically formulate all the possible DSs that $\vec{E}$ may exhibit by considering all possible products of the above building block operators (on both scales). This combinatorial large ensemble of options may be slightly reduced in size by considering that: (i) for the group to exhibit closure, a DS can only combine different operations that have a commensurate order [4]; and (ii) we are only interested in the new types of DSs that involve operations on the macroscopic scale (as the microscopic theory is already well-established). With this in mind, we begin mapping out DSs according to their dimensionality and whether or not they involve different length scales. For instance, we define a general operation that involves both a temporal operator (e.g., time translation or reversal) coupled to a macroscopic operation (e.g., rotation or reflection), hence denoted as macrospace-time operation, $\widehat{M}$, by:

$$\widehat{M}\vec{E}(\vec{R},t) = \vec{E}(\hat{\Gamma}_R \vec{R} + \vec{u}, st + \tau) = \vec{E}(\hat{\Gamma}\vec{X} + \vec{a}) \qquad (2)$$

where $\hat{\Gamma}_R$ is an $M$-dimensional point group operation, $s = \pm 1$ indicates the resulting possible action of time-reversal, and $(\vec{u}, \tau) \equiv \vec{a}$ is a vector that denotes translations in space and time, respectively. Table 1 in the SI summarizes the macrospace-time operation building blocks, which construct the general operator ($\widehat{M}$). We note that the order of the composite operation is the lowest common multiple of the order of the building block operators.

It is instructive to consider a concrete physical example of an EM field that exhibits such a macrospace-time symmetry. For instance, an EM field with temporal frequency, $\omega$, carrying orbital angular momentum (OAM) which is characterized by the phase winding number (or topological charge), $l$, has a continuous symmetry of macroscopic rotation with time translation: $\vec{E}(\theta,t) = \vec{E}(\theta + \delta, t + l\delta/\omega)$, i.e., it is invariant under $\widehat{M} = \hat{R}_\delta \hat{\tau}_{l\delta}$ (Fig. 1a), where $\delta$ is any real number. Figure 1b shows another example of the macrospace-time symmetry of a superposition of two fields at frequencies $\omega_2 = 3\omega_1$ and angular momenta $l_1 = 1$ and $l_2 = $ -1, which exhibit the symmetry, $\vec{E}(\theta,t) = \vec{E}(\theta - 2\pi/4, t + T/4)$, i.e., $\widehat{M} = \hat{R}_{4,-1}\hat{\tau}_{4,1}$. That is, the overall EM field is invariant under the combined operations of a rotation by $-2\pi/4$ in $\theta$ and a time translation by $T/4$.

Next, we explore multi-scale DSs (i.e. products of both microscopic and macroscopic operations, which are typically different from each other). These operations take on the following form:

$$\hat{\gamma}\widehat{M}\vec{E} = \hat{\gamma}\vec{E}(\hat{\Gamma}\vec{X} + \vec{a}) \qquad (3)$$

where $\hat{\gamma}$ is a point-group microscopic operation that acts on the polarization DOF of $\vec{E}$. For a general vector field, $\hat{\gamma}$ can act on all local parameters like spin [28] and color [29,30]. Table 2 in the SI summarizes all the building blocks for microscopic operation. As an example, Fig. 1c shows a superposition of three twisted EM beams with circular polarization and the high-order multi-scale DS, $\hat{R}_{5,-1}\hat{\tau}_{5,1}\hat{r}_{5,2}$. That is, the overall EM field in Fig. 1c is invariant under the following combined operations: a rotation by $-2\pi/5$ in $\theta$, a time translation by $T/5$, and a rotation of the polarization by $4\pi/5$. This is an example of a DS that combines macroscopic and microscopic operations, where the field itself is NOT invariant under only microscopic or macroscopic operations. The theory describes the physical manifestation of coupling between the different DOF of the EM field.

EM fields usually exhibit many DSs. For instance, the field in Fig. 1b possesses both $\hat{R}_{4,-1}\hat{\tau}_{4,1}$ and $\hat{\imath}\hat{R}_{4,1}\hat{\tau}_{4,1}$ along with their various products and powers. The combination of DSs is best described by a group theory where each group is closed and formed by a finite set of generating DSs. With this approach, the complete EM field can be obtained from the field information within a single unit cell (which can be smaller than [0,$T$] and [0,$\vec{L}$] temporal/spatial periods of the EM field) and the comprising symmetry group (similarly to solid state lattices).

**DSs of the induced polarization**

Having discussed in general the DSs that characterize EM fields, we go on to apply the symmetry theory to light-matter interactions, focusing as an example on HHG. We consider the nonlinear interaction of a macroscopic medium irradiated by an electric field. Within the Born-Oppenheimer and dipole approximations, the *microscopic* Hamiltonian of a nonlinear system at a *macroscopic* point, $\vec{R}$, interacting with a laser field is given in atomic units and in the length gauge by:

$$\hat{H}_{\vec{R}}(t) = \sum_j -\frac{\nabla_j^2}{2} + \frac{1}{2}\sum_{i\neq j}|\vec{r}_i - \vec{r}_j|^{-1} + \sum_j U_{\vec{R}}(\vec{r}_j) + \sum_j \vec{E}(\vec{R},t)\cdot\vec{r}_j \qquad (4)$$

where $\vec{r}_j$ is the *microscopic* coordinate of the $j$th electron, and $U_{\vec{R}}$ is the time-independent potential that is associated with the electrostatic interactions of electrons in the system with the nuclei. We assume here that the long-range interactions between the different *macroscopic* points is negligible; therefore; the full wave function of the non-interacting microscopic systems is a non-interacting product of the microscopic wavefunctions in different spatial positions: $|\psi(t)\rangle = \prod_{\vec{R}}|\psi_{\vec{R}}(t)\rangle$, where $|\psi_{\vec{R}}(t)\rangle$ is the wavefunction of a microscopic system located at $\vec{R}$. Even if the microscopic Hamiltonian and wave function lack any microscopic DS, the total Hamiltonian that includes the macroscopic structure, $\hat{H}(t) = \sum_{\vec{R}}\hat{H}_{\vec{R}}(t)$, and the full wavefunction can have multi-scale DSs. This can happen if: (i) the electric field $\vec{E}(\vec{R},t)$, (ii) the sum of the microscopic potentials $U = \sum_{\vec{R}} U_{\vec{R}}$, and (iii) and the wave function of the initial state (which is typically the ground state of $\hat{H}(t)$ with $\vec{E}(\vec{R},t) = 0$, however, it can also be an excited state [31,32]) all exhibit a shared multi-scale DS (for proof see next section). When $|\psi(t)\rangle$ exhibits a DS, any measured observable, $o(\vec{R},t) = o(\vec{R},t) = \langle\psi(t)|\hat{O}|\psi(t)\rangle = \langle\psi(t)|\sum_{\vec{R}}\hat{O}_{\vec{R}}|\psi(t)\rangle$ (e.g., the induced polarization, $\vec{P}(\vec{R},t)$), also upholds that DS.

**Constraints on the Fourier spectrum of the induced polarization**

To analytically derive the selection rules due to DSs exhibited by $\hat{H}(t)$, we analyze a general $\vec{P}(\vec{R},t)$ function in the Fourier-domain:

$$\vec{P}(\vec{R},t) = \vec{P}(\vec{X}) = \sum_{\vec{k}} \vec{F}(\vec{k})\exp(i\vec{k}\cdot\vec{X}) \qquad (5)$$

where $\vec{k} = \sum_{j=1}^{\widetilde{D}} q_j\vec{b}_j$. Here, $\{\vec{b}_j\}$ is a set of $\widetilde{D}$, the total macroscopic and time dimensionality, independent vectors that comprise the temporal and spatial frequencies, and $q_j$ are integers which represent the harmonic order of the $\vec{b}_j$ frequencies. The Fourier coefficients, $\vec{F}(\vec{k})$, are N-dimensional complex-valued vectors. It is useful to express $\vec{F}(\vec{k})$ as $\vec{F}(\vec{k}) = A(\vec{k})\exp(i\phi(\vec{k}))\hat{F}(\vec{k})$, where $A(\vec{k})$ is a real positive amplitude, $\phi(\vec{k})$ is a real number, and $\hat{F}(\vec{k})$ is the unit vector, i.e., the polarization direction of $\vec{F}(\vec{k})$. Since $\vec{F}(\vec{k})$ denotes the Fourier decomposition of $\vec{P}(\vec{X})$, any DS exhibited by $\vec{P}(\vec{X})$ leads directly to restrictions, i.e., selection rules, on the spectral behavior of $\vec{F}(\vec{k})$. The three different kinds of restrictions are: (i) forbidden harmonics, (ii) forbidden polarization of some harmonics, and (iii) harmonics that must have the same amplitude and polarization up to reflection or rotations. Importantly, 'harmonics' in the context of selection rules mean both the temporal and spatial frequency content – that is, selection rules can, and often do, lead to mixed restrictions that couple the spatial frequencies and angular momenta with the temporal frequency and polarization components. In that sense, the multi-scale theory leads to new types of generalized restrictions linking all DOF of the emitted light.

Next, we derive the selection rules. When $\vec{P}(\vec{X})$ is invariant under some multi-scale DS:

$$\vec{P}(\vec{X}) = \hat{G}^{-1}\vec{P}(\vec{X}) = \hat{\gamma}^{-1}\vec{P}(\hat{M}\vec{X}) = \hat{\gamma}^{-1}\vec{P}(\hat{\Gamma}^{-1}\vec{X} - \vec{a}) \tag{6}$$

, its Fourier decomposition must fulfill (for full derivation see SI):

$$\hat{\gamma}\vec{F}(\hat{\Gamma}\vec{k})\exp(i\vec{k}\cdot\vec{a}) = \vec{F}(\vec{k}) \tag{7}$$

Eq. (7) states that only certain harmonics are allowed, i.e., the symmetry manifests in the spectral structure of $\vec{F}(\vec{k})$. Consequently, eq. (7) can be written as

$$A(\hat{\Gamma}\vec{k})\exp(i\phi(\hat{\Gamma}\vec{k}) + i\vec{k}\cdot\vec{a})\hat{\gamma}\hat{F}(\hat{\Gamma}\vec{k}) = A(\vec{k})\exp(i\phi(\vec{k}))\hat{F}(\vec{k}) \tag{8}$$

, which has a non-trivial solution when all of the following three conditions hold simultaneously: (i) The amplitude of the $\vec{k}$ harmonic is equal to that of the $\hat{\Gamma}\vec{k}$ harmonic, i.e. $A(\vec{k}) = A(\hat{\Gamma}\vec{k})$. (ii) The polarization of the $\hat{\Gamma}\vec{k}$ harmonic under a $\hat{\gamma}$ microscopic operation is equal to the polarization of the $\vec{k}$ harmonic up to some $\alpha_i$ phase, i.e., $\hat{\gamma}\hat{F}(\hat{\Gamma}\vec{k}) = \exp(i\alpha_i)\hat{F}(\vec{k})$. This is possible if the microscopic operation, $\hat{\gamma}$, is a product of two microscopic operations, $\widehat{\gamma'}$ and $\widehat{\gamma''}$, such that $\hat{\gamma}\hat{F}(\vec{k}) = \widehat{\gamma''}\widehat{\gamma'}\hat{F}(\vec{k}) = \exp(i\alpha_i)\widehat{\gamma''}\hat{F}_i(\vec{k})$, i.e., $\hat{F}_i(\vec{k})$ and $\exp(i\alpha_i)$ are the $i$th eigenvector and eigenvalue of $\widehat{\gamma'}$ (Table 2 in the SI lists the possible eigenvectors and eigenvalues), respectively, and $\hat{F}(\hat{\Gamma}\vec{k}) = \widehat{\gamma''}^{-1}\hat{F}(\vec{k})$). (iii) The phase $\phi_i(\vec{k})$ uphold the equation:

$$\phi_i(\vec{k}) = \vec{k}\cdot\vec{a} + \alpha_i + \phi_i(\hat{\Gamma}\vec{k}) - 2\pi Q \tag{9}$$

where $Q$ is an integer.

**Selection rules**

Eq. (9), specifically, has different kinds of solutions depending on $\hat{\Gamma}$. For instance, when $\hat{\Gamma}$ is the identity operator, then the selection rule has the form, $\vec{k}\cdot\vec{a} + \alpha_i = 2\pi Q$. This form appears in the selection rules in rows 4-8 in table 1. This type of solution describes the allowed harmonics and their polarization in a similar manner to that of microscopic HHG. Alternatively, when the DS also involves macroscopic rotations ($\hat{\Gamma} = \hat{R}_{n,m}$), the equation for the phase of $\vec{F}(\vec{k})$ becomes $\phi_i(\hat{R}_{n,m}\vec{k}) = \phi_i(\vec{k}) + 2\pi lm/n$. Then eq. (9) leads to the condition $\vec{k}\cdot\vec{a} + 2\pi lm/n + \alpha_i = 2\pi Q$. Here, $l$ is the allowed phase winding number that characterizes the OAM of the emitted $\vec{k}$ harmonic. So the selection rules in this case would describe the allowed harmonic indices, their OAMs, and polarization states. When the DS involves macroscopic reflections or time-reversal (e.g. $\hat{\Sigma}_x$ or $\hat{T}$), this restricts a pair of harmonics, e.g., $\vec{k}$ and $\hat{\Sigma}_x\vec{k}$, to have to same amplitude. The relationship between the polarization of each harmonic in the pair then depends on the microscopic operation, $\hat{\gamma}$. When $\hat{\gamma}$ is the identity operator or a rotation, the pair of harmonics have identical polarizations. When $\hat{\gamma}$ is a improper-rotation (e.g. reflection), the polarizations of the harmonic pair are reflections of each other. In general case, all these different DOF may be coupled. In the SI we derive the selection rules for symmetries that include time reversal or space reflection (rows 9-14 in table 1, for the case of (1+1+2)D).

Up until now, we have discussed discrete symmetries that are combinations of discrete operations. However, $\hat{G}$ can also be a continuous operator. For example the continuous DSs $\hat{r}_{\delta\alpha}\vec{E}(\hat{R}_{\delta(n,m)}\vec{X} + \delta\vec{a}) = \vec{E}(\vec{X})$ for any real $\delta$, leads according to eq.(9) to the selection rules:

$$\vec{k} \cdot \vec{a} + 2\pi lm/n \pm \alpha = 0, \qquad (10)$$

which means that the combination of polarization, energy, and linear and angular momenta is constant and can be considered a conserved charge, in a similar manner to the conservation of torus-knot angular momentum described in [6].

Table 1 summarizes the different DSs in the (2+1+1)D case (i.e., where there are two microscopic polarization dimensions, one time axis, and one macroscopic axis) and their associated selection rules. We again emphasize that these involve new selection rules as compared to the microscopic theory [4], i.e., the inclusion of macroscopic DOF in the EM field can change the system's response and lead to new control mechanisms for XUV light in HHG, or new routes for ultrafast spectroscopy. Higher dimensionalities can be similarly derived.

**Table 1**- DSs and their associated selection rules in the (2+1+1)D case (i.e. where there are two microscopic polarization dimensions, one time axis, and one macroscopic axis). The harmonic order of the temporal (spatial) frequency is $q_1$ ($q_2$). Notably, the selection rules are affected by the inclusion of macroscopic operations. In row 1, a continuous space-time translation DS leads to energy-momentum conservation [16,33]. In row 2, a continuous space-time translation and microscopic rotation DS leads to energy-momentum-spin conservation [5,6,17]. In row 3, the DS is the same as in row 2, except now the rotation is along an ellipse in which the ratio of the minor to major axis is $b$. This leads to energy-momentum-ellipticity conservation.

|    | Dynamical symmetry | Spectral selection rule | Polarization |
|----|---|---|---|
| 1  | $\hat{\tau}(\delta)\hat{J}(\beta\delta)$ | $q_1 + \beta q_2 = 0$ | Any |
| 2  | $\hat{\tau}(\delta)\hat{J}(\beta\delta)\hat{r}(\alpha\delta)$ | $q_1 + \beta q_2 \pm \alpha = 0$ | ($\pm$) Circular |
| 3  | $\hat{\tau}(\delta)\hat{J}(\beta\delta)\hat{e}(\alpha\delta)$ | $q_1 + \beta q_2 \pm \alpha = 0$ | Ellipticity of $\pm b$ |
| 4  | $\hat{\tau}_2\hat{J}_2, \hat{\tau}_2\hat{J}_2\hat{r}_2$ | $q_1 + q_2 = odd$ | Any |
| 5  | $\hat{\tau}_2\hat{J}_2\hat{\sigma}$ | $q_1 + q_2 = even$ | In $\hat{\sigma}$ plane |
|    |  | $q_1 + q_2 = odd$ | Orthogonal to $\hat{\sigma}$ plane |
| 6  | $\hat{\tau}_{n_0}\hat{J}_{n_0,m}$ | $q_1 + q_2 = n_0 n$ | Any |
| 7  | $\hat{\tau}_{n_0,m_1}\hat{J}_{n_0,m_2}\hat{r}_{n_0,m_3}$ | $q_1 m_1 + q_2 m_2 \pm m_3 = n_0 n$ | ($\pm$) Circular |
| 8  | $\hat{\tau}_{n_0,m_1}\hat{J}_{n_0,m_2}\hat{e}_{n_0,m_3}$ | $q_1 m_1 + q_2 m_2 \pm m_3 = n_0 n$ | Ellipticity of $\pm b$ |
| 9  | $\hat{I}, \hat{I}\hat{r}_2, \hat{I}\hat{\tau}_2\hat{J}_2\hat{r}_2$ | $A_{q_1,q_2} = A_{-q_1,-q_2}$ | Linear, $\hat{F}_{q_1,q_2} = \hat{F}_{-q_1,-q_2}$ |
| 10 | $\hat{I}\hat{\sigma}, \hat{I}\hat{\tau}_2\hat{J}_2\hat{\sigma}$ | $A_{q_1,q_2} = A_{-q_1,-q_2}$ | Ellipse axis orthogonal to $\hat{\sigma}$ plane, $\hat{F}_{q_1,q_2} = \hat{\sigma}\hat{F}_{-q_1,-q_2}$ |
| 11 | $\hat{\Sigma}_x\hat{\tau}_2, \hat{\Sigma}_x\hat{\tau}_2\hat{r}_2$ | $A_{q_1,q_2} = A_{q_1,-q_2}$ | $\hat{F}_{q_1,q_2} = \hat{F}_{q_1,-q_2}$ |
| 12 | $\hat{\Sigma}_x\hat{\tau}_2\hat{\sigma}$ | $A_{q_1,q_2} = A_{q_1,-q_2}$ | $\hat{F}_{q_1,q_2} = \hat{\sigma}\hat{F}_{q_1,-q_2}$ |
| 13 | $\hat{T}\hat{J}_2, \hat{T}\hat{J}_2\hat{r}_2$ | $A_{q_1,q_2} = A_{-q_1,q_2}$ | Linear |
| 14 | $\hat{T}\hat{J}_2\hat{\sigma}$ | $A_{q_1,q_2} = A_{-q_1,q_2}$ | Ellipse axis orthogonal to $\hat{\sigma}$ plane |

## DSs and selection rules of twisted light

We demonstrate numerically that these analytically derived selection rules are upheld by using the Lewenstein model [34] to calculate the time-dependent dipole $\vec{P}(\vec{R},t)$ generated by the field $\vec{E}(\vec{R},t)$. An example is shown in Fig. 1, $\vec{E}(\theta,t) \propto \hat{e}_+ cos(2\omega t - \theta) + \hat{e}_-[cos(2\omega t) + cos(3\omega t + \theta)]$, where $\hat{e}_+$ and $\hat{e}_-$ are the left- and right-rotating circularly polarized polarizations, respectively. This field exhibits a DS of $\hat{\tau}_{5,1}\hat{R}_{5,-1}\hat{r}_{5,2}$, and therefore, the selection rule for the harmonics emitted from this field (derived in Table 1 row 7) is $q - l \pm 2 = 5n$. Here $q$ is the temporal harmonic, $l$ is the OAM winding number, and $n$ is an integer, such that this selection rule essentially couples the harmonic order and its angular momenta and polarization in one generalized constraint, i.e. certain harmonics orders can only be emitted with certain values of OAM with ($\pm$) circular polarization. This type of coupling is not possible without the multi-scale operations. The selection rule is numerically investigated in Fig. 1d, which shows the intensity of the left- and right-rotating harmonics as a function of $q$ and $l$, agreeing well with the analytic theory.

Our theory is also applicable for non-isotropic and inhomogeneous media. For example, following a recent publication [35], we investigate the symmetries and selection rules of a circularly polarized field with orbital momentum, $\vec{E} \propto \hat{e}_+ \exp(i(\omega t + l_0 \theta))$, propagating along the optical axis of a β-BaB2O4 (BBO) crystal (Fig. 1e). Ref [35] explored experimentally the emitted SHG in this system, with $l_0$ equals 0 or 1, and observed that SHG fields of the form: $\hat{e}_- \exp(i(2\omega t + 2l_0\theta))$, $\hat{e}_+ \exp(i(2\omega t + (2l_0 - 2)\theta))$, and $\hat{e}_+ \exp(i(2\omega t + (2l_0 + 4)\theta))$, were generated. These results were explained by a cascaded linear spin–orbit interaction and a perturbative nonlinear interaction. As shown below, our analysis directly yields all the allowed channels, only some of which were obtained by the cascaded model [35].

The BBO crystal irradiated by the field $E \propto \hat{e}_+ \exp(i(\omega t + l_0 \theta))$ have the combined symmetry of $\hat{\tau}_{6,l_0+4}\hat{R}_{6,-1}\hat{r}_{6,2}$. Therefore, according to eq.(9), the selection rules are: $q(l_0 + 4) - l + 2s = 6Q$, where $q$ is the temporal harmonic, $l$ is the OAM winding number, $s$ is the spin, and Q is an integer. Hence, for SHG ($q = 2$), we get the selection rule for the OAM with spin, $s = \pm 1$:

$$l = 6Q + 2 + 2l_0 \pm 2 \tag{11}$$

This result agrees with measured and predicted results in [35], and also predict more allowed channels. Figure 1f presents the rich structure of the allowed channels, even though the setup is simple.

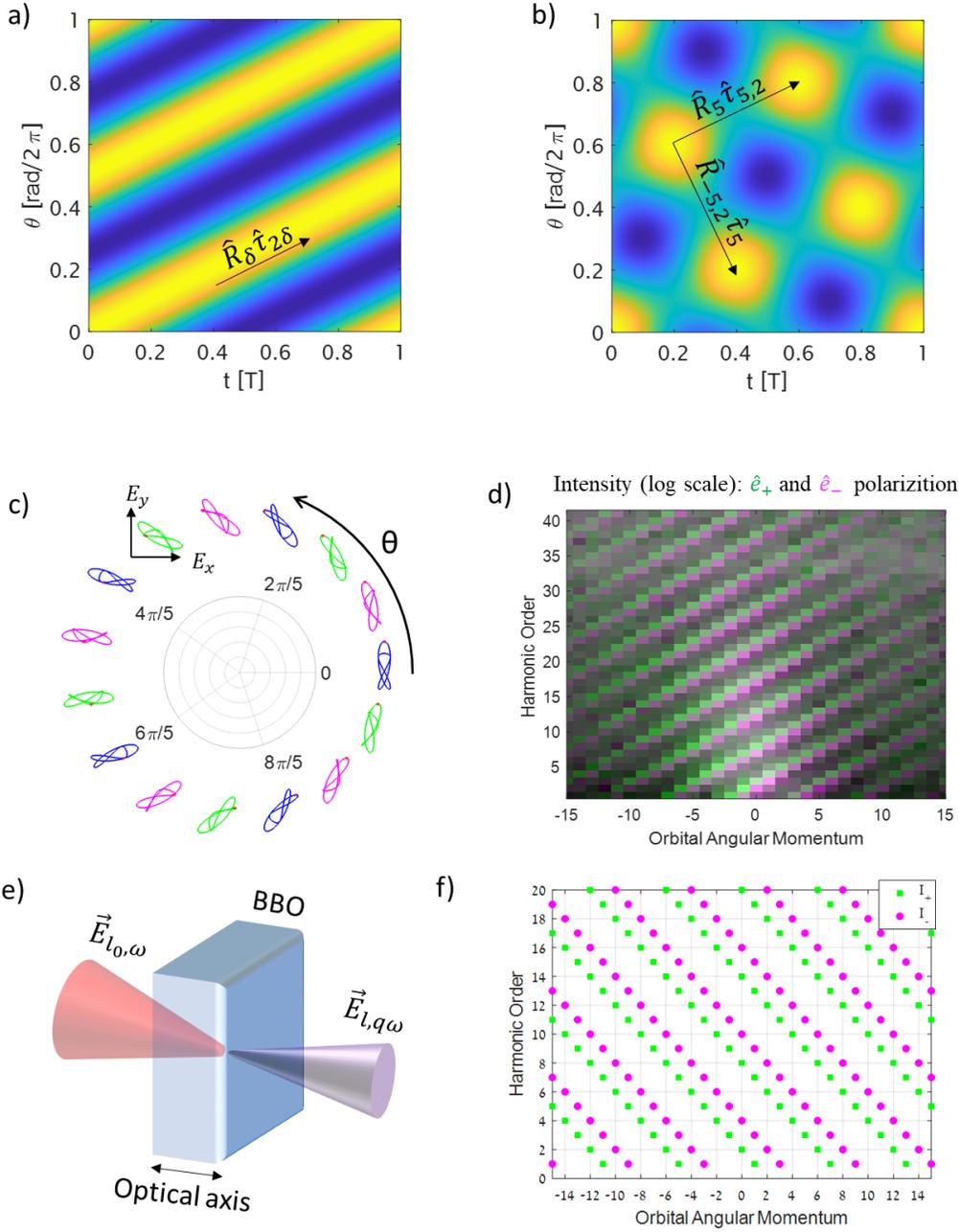

**Fig. 1.** a) A monochromatic field with OAM ($\vec{E}(\theta,t) \propto \cos(\omega t + 2\theta)$) with a continuous symmetry of macroscopic rotation and time translation, $\vec{E}(\theta,t) = \vec{E}(\theta + \delta, t + l\delta/\omega)$, i.e., $\hat{R}_\delta \hat{\tau}_{l\delta}$. b) The superposition of two fields, $\cos(\omega t + \theta) + \sin(3\omega t - \theta)$, has the fourth order discrete symmetries, $\hat{R}_{4,-1}\hat{\tau}_{4,1}$ and $\hat{\imath}\hat{R}_{4,1}\hat{\tau}_{4,1}$. c) The superposition of three circularly polarized fields, $\hat{e}_+ \cos(2\omega t - \theta) + \hat{e}_-[\cos(2\omega t) + \cos(3\omega t + \theta)]$, exhibits the multi-scale DS, $\hat{R}_{5,-1}\hat{\tau}_{5,1}\hat{r}_{5,2}$. The Lissajous curve of the local field is plotted every $2\pi/15$ rotation of $\theta$. The shape of the Lissajous curve repeats itself every $2\pi/5$ rotation in $\theta$ followed with rotation of the polarization by $-4\pi/5$ and $-T/5$ time translation which is indicated by a red dot at $t = 0$. However, there is no symmetry between the local fields at arbitrary angles, as can be seen in the non-identical Lissajous curves for fields that are separated by $2\pi/15$. d) The simulated intensity of the harmonics emitted from the field in c) clearly demonstrates the selection rule, $q - l \pm 2 = 5Q$, for left ($\hat{e}_+$) and right ($\hat{e}_-$) circularly polarized harmonics. e) Circularly polarized field with orbital momentum, $\vec{E} \propto \hat{e}_+ \exp(i(\omega t + l_0 \theta))$, propagating along the optical axis of a β-BaB2O4 (BBO) crystal generating harmonic field. f) The selection rules for the harmonic order, OAM and helicity of the generated harmonics in the BBO crystal.

## Space-time polarized quasicrystal

We now apply the theory to non-periodic systems, e.g., quasicrystals [25] and optical quasicrystals [26,27], by employing the superspace concept [23] where $M$ (the effective number of macroscopic DOF) is larger than the physical macroscopic dimension. We demonstrate this option by deriving the HHG selection rules driven by a field with a spatiotemporal quasi-periodic structure. Specifically, we consider a superposition of four fields that interact with a thin medium. The field, limited to the $(t, X)$ axes (i.e., dimensionality of (2+1+1)D), is described by:

$$\vec{E}(t, X) = A\left[\cos(\omega t + (1+\sqrt{2})kX) - \cos\left((1+\sqrt{2})\omega t - kX\right)\right]\hat{x} \\ + \left[\cos\left((1+\sqrt{2})\omega t + kX\right) - \cos(\omega t - (1+\sqrt{2})kX)\right]\hat{y} \quad (12)$$

This field has a vectorial quasicrystal structure in space-time, as shown in Fig. 2a. Utilizing the superspace representation (with dimensionality (2+4+4)D), the field can be re-written in superspace as:

$$\vec{E}(t_1, t_2, t_3, t_4, X_1, X_2, X_3, X_4) \\ = A\left[\cos(\omega t_1 + (1+\sqrt{2})kX_1) - \cos\left((1+\sqrt{2})\omega t_2 - kX_2\right)\right]\hat{x} \\ + \left[\cos\left((1+\sqrt{2})\omega t_3 + kX_3\right) - \cos(\omega t_4 - (1+\sqrt{2})kX_4)\right]\hat{y} \quad (13)$$

Where $X_j$ are new coordinates introduced in the higher dimensionality of the superspace representation. In this representation, $\vec{E}$ is periodic in all coordinates $X_{j=1,2,3,4}$. Therefore, also the induced polarization will be periodic in $X_{j=1,2,3,4}$. Hence, the induced polarization can be expressed with spatial frequencies of $\vec{k} = \sum_{j=1}^{4} q_j \vec{b}_j = q_1(1+\sqrt{2})k\hat{X}_1 - q_2 k\hat{X}_2 + q_3 k\hat{X}_3 - q_4(1+\sqrt{2})k\hat{X}_4$, where $q_j$ is an integer, which also corresponds to the number of photons annihilated from each field.

In order to obtain the HHG selection rules, at this point we project the superspace representation back to physical space, i.e. take $\hat{X}_1 = \hat{X}_2 = \hat{X}_3 = \hat{X}_4 \to \hat{X}$. Following this, the allowed $X$-axis spatial frequencies are $k_{\vec{q}}^{(X)} = q_1(1+\sqrt{2})k - q_2 k + q_3 k - q_4(1+\sqrt{2})k$, which is equivalent to a condition of conservation of $X$-axis momentum. Other symmetries also connect the spatial and temporal harmonic orders. For instance, the continuous symmetry $\vec{E}\left(t_1 + \frac{\delta}{\omega}, t_2, t_3, t_4, X_1 - \delta/(1+\sqrt{2})k, X_2, X_3, X_4\right) = \vec{E}(\vec{t}, X_1, X_2, X_3, X_4)$ for any $\delta$, which is associated with the evolution of only the first subfield of the four subfields superposition in eq. (12). By applying eq. (9), we arrive at $\vec{k} \cdot \vec{a} = q_1' - q_1 = 0$, where $q_1$ is the spatial $\hat{X}_1$ harmonic and $q_1'$ is the temporal $t_1$ harmonic. Therefore, $q_1' = q_1$. Similarly this relationship is valid to all the subfields, hence $q_j' = q_j$, indicating that the emitted $\omega_{\vec{q}}$ temporal harmonics have the same $q_j$ of $k_{\vec{q}}^{(X)}$: $\omega_{\vec{q}} = q_1 \omega + q_2(1+\sqrt{2})\omega + q_3(1+\sqrt{2})\omega + q_4\omega$. In the photonic picture, this means that each annihilated photon gives its energy and momentum to the generated $\omega_{\vec{q}}$ photon.

The field in eq. 12 also exhibits the following two DSs: $\hat{\sigma}_x \vec{E}\left(t, X_1 + \frac{\pi}{(1+\sqrt{2})k}, X_2 + \frac{\pi}{k}, X_3, X_4\right) = \vec{E}(t, X_1, X_2, X_3, X_4)$ and $\hat{\sigma}_y \vec{E}\left(t, X_1, X_2, X_3 + \frac{\pi}{k}, X_4 + \frac{\pi}{(1+\sqrt{2})k}\right) = \vec{E}(x, t_1, t_2, t_3, t_4)$. Applying our theory to these DSs (eq. 7) yields:

$$\hat{\sigma}_x \vec{F}(\vec{k}) \exp(i(q_1 + q_2)\pi) = \vec{F}(\vec{k}) \\ \hat{\sigma}_y \vec{F}(\vec{k}) \exp(i(q_3 + q_4)\pi) = \vec{F}(\vec{k}) \quad (14)$$

Eq. 14 dictates that harmonics with an odd $q_1 + q_2$ and an even $q_3 + q_4$ are x-polarized, harmonics with an even $q_1 + q_2$ and an odd $q_3 + q_4$ are y-polarized, and all other harmonics are forbidden. The field in eq. (12) also exhibits higher symmetries due to its space-time polarized octagonal quasicrystal structure. The octagonal quasicrystal with an eight-fold symmetry is well-known in crystallography of two-dimensional static spatial arrangements of atoms [36]. Here the vector electric field exhibits symmetries that are combinations of the $d_8$ dihedral group operations in space-time followed by microscopic operations:

$$\hat{r}_4 \vec{E}\left(\hat{R}_8(\zeta,\eta)\right) = \hat{\sigma}_{\frac{3\pi}{4}} \vec{E}\left(\hat{\Sigma}_{0\ or\frac{\pi}{2}}(\zeta,\eta)\right) = \hat{\sigma}_{\frac{\pi}{4}} \vec{E}\left(\hat{\Sigma}_{\frac{\pi}{4}\ or\frac{3\pi}{4}}(\zeta,\eta)\right) = \vec{E}\left(\hat{\Sigma}_{\frac{1+2n}{8}\pi}(\zeta,\eta)\right) = \vec{E}(\zeta,\eta) \quad (15)$$

where the subscript of $\hat{\sigma}$ ($\hat{\Sigma}$) denote the angle of the reflection axis in the $x - y$ ($\zeta - \eta$) plane, and $n$ is an integer, and where $\zeta = kx$ and $\eta = \omega t$ are the dimensionless space-time variables. According to eq. (7), the polarized octagonal symmetries appear also in the Fourier domain of the driving field in eq. (12). This field has eight peaks in the Fourier domain (fig. 2b), where each peak is a linearly polarized along $\hat{x}$ or $\hat{y}$ rotated $\pi/2$ from the next peak, which is rotated by $\pi/4$ in the Fourier plane, as expected from eq. (15). In addition, the induced polarization exhibits symmetries in the space-time domain (fig. 2c) and in the Fourier domain (fig. 2d):

$$\hat{r}_4 \vec{F}\left(\hat{R}_8 \vec{k}\right) = \hat{\sigma}_{\frac{3\pi}{4}} \vec{F}\left(\hat{\Sigma}_{0\ or\frac{\pi}{2}} \vec{k}\right) = \hat{\sigma}_{\frac{\pi}{4}} \vec{F}\left(\hat{\Sigma}_{\frac{\pi}{4}\ or\frac{3\pi}{4}} \vec{k}\right) = \vec{F}\left(\hat{\Sigma}_{\frac{1+2n}{8}\pi} \vec{k}\right) = \vec{F}(\vec{k}) \quad (16)$$

where $\vec{k}$ is a vector composed of the two dimensionless frequencies, $\zeta$ and $\eta$.

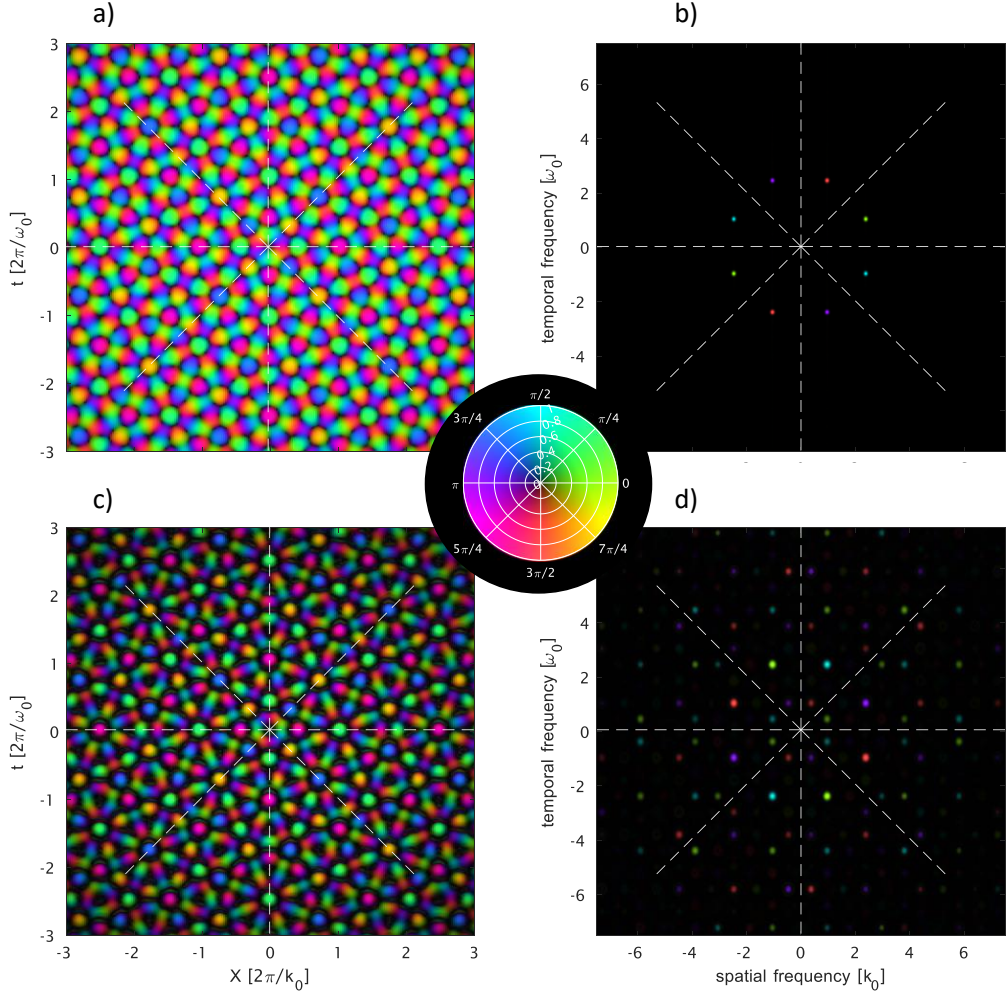

**Fig. 2.** Nonlinear wave mixing in optical time-space vectorial quasicrystal. a) The electric field in normalized space and time poses an octagonal quasicrystal vectorial structure. The direction and amplitude of the polarization is mapped to a color according to the center plot. b) The Fourier representation of the octagonal quasicrystal field has a simpler octagonal structure with eight peaks, where the polarization from one peak to the other is rotated by $\pi/2$. c) The complex quasicrystal symmetry is conserved in the simulated induced polarization, $\vec{P}$. d) More peaks, i.e., more harmonics, are generated in the induced polarization, and the polarized octagonal structure is conserved.

**Experimental multi-scale DS**

We investigate experimentally an HHG selection rule that is based on a (2+1+1)D multi-scale DS, i.e. DS in two microscopic dimensions, time and one macroscopic dimension which is the propagation axis. The experimental laser field consists of three beams: a bi-circular beam and additional Bessel beam, with the following form:

$$\vec{E}_\varphi(t,z,\rho) = \hat{e}_+ e^{-(\rho/w)^2} a_1 \cos(\omega t') + \hat{e}_- e^{-(2\rho/w)^2} a_2 \cos(2\omega t') + \vec{E}^{Bessel}{}_\varphi(t,z,\rho) \quad (17)$$

where $z$ is the propagation axis, $\rho$ is the radial axis, $t'$ is the retarded time ($t' = t - z/c$), $w$ is the waist of the $\omega$ Gaussian beam, and $a_1$ and $a_2$ are the beams amplitudes. The Bessel beam is given by:

$$\vec{E}^{Bessel}{}_\varphi(t,z,\rho) = a_3(\sin(\varphi + \pi/4)\hat{e}_+ + \sin(\varphi - \pi/4)\hat{e}_-)\cos(\omega t' + \beta z)J_0(\sqrt{2k\beta - \beta^2}\rho) \quad (18)$$

where $a_3$ is the amplitude, $k = \omega/c$ is the wave vector, $\beta$ is the on-axis difference between the wave vectors of the Gauss and Bessel beams, $J_0$ is the zero order Bessel function of the first kind, and $\varphi$ controls the polarization of the Bessel beam (In this experiment, $\varphi$ corresponds to the angle between the optical axis of the quarter waveplate (QWP) and the axis of the incoming linearly polarized Bessel beam). The use of the Bessel beam allows us to have tunable control over the shape and structure of the light beam in mutli-scale dimensionality - the polarization states of the beams control the microscopic dimensions, and the difference phase velocity between the bi-circular and Bessel beams allows to generate macroscopic structures along the z-axis (the beams propagation axis). Overall, these knobs allow us to tune the DSs of the field, including generation of multi-scale symmetries, even when it does not have a unique symmetry in just microscopic space. For example, when $\varphi = \pi/4$, the polarization of $\vec{E}^{Bessel}$ is $\hat{e}_+$, similar to that of the $\omega$ Gaussian beam. Therefore, the total field has the microscopic DS, $\hat{r}_{3,1}\vec{E}_{\varphi=\frac{\pi}{4}}(t - T/3, z, \rho) = \vec{E}_{\varphi=\frac{\pi}{4}}(t, z, \rho)$, i.e., the same DS of the bi-circular field that leads to the selection rule, $q = 3Q \pm 1$, with circular polarization, $\hat{e}_\pm$ [37]. On the other hand, for $\varphi = -\pi/4$ (and a polarization of $\vec{E}^{Bessel}_{\varphi=-\pi/4}$ that is $\hat{e}_-$, like the $2\omega$ Gaussian beam), the total field has no microscopic DS; rather, it has the multi-scale DS, $\hat{r}_{3,1}\vec{E}_{\varphi=-\frac{\pi}{4}}(t - T/3, z + L/3, \rho) = \vec{E}_{\varphi=-\frac{\pi}{4}}(t, z, \rho)$, where $L = 2\pi/\beta$ is the spatial period of the total field. According to eq. (9), this DS should lead to the selection rule, $q_1 - q_2 = 3Q \pm 1$, with circular polarization, $\hat{e}_\pm$, where $q_1$ and $q_2$ are the temporal and longitudinal harmonic orders, respectively. Finally, when $\varphi \neq \pi/4 + n\pi$, the total field lacks any DS, and therefore all the harmonic orders and polarizations are allowed.

We shall now focus on the $\varphi = -\pi/4$ case that leads to multi-scale DS. Figure 3a emphasizes the $\hat{Z}_{3,1}\hat{t}_{3,-1}\hat{r}_{3,1}$ DS of the $\vec{E}_{\varphi=-\frac{\pi}{4}}$ field, where the time domain Lissajou curve is plotted every $z$ step which visualizes a 3D surface (color-coding represents time). The three blue Lissajou plots are separated by $L/3$ from each other and show the multi-scale $\hat{Z}_{3,1}\hat{t}_{3,-1}\hat{r}_{3,1}$ DS. Notably, this is a symmetry of all the electric fields involved in the light-matter interaction: each component of $\vec{E}$ and the XUV field (Fig. 3b). In the experiment, we only measured the XUV emission with small divergence, i.e., with $q_2 = 0$. This on-axis emitted XUV field is summed up coherently along the z-axis during propagation (Fig. 3b). We first explore HHG in this scheme numerically. Figure 3c presents the microscopic yield of different harmonics as a function of $\varphi$. As shown, only $\varphi = \pi/4$ (with the microscopic $\hat{t}_{3,-1}\hat{r}_{3,1}$ DS) leads to the forbidden $3Q$ harmonics. For $\varphi = -\pi/4$, the yield of the $3Q$ harmonics is maximized. When we include propagation (Fig. 3d), the $3Q$ harmonics are also forbidden at $\varphi = -\pi/4$, due to the multi-scale $\hat{Z}_{3,1}\hat{t}_{3,-1}\hat{r}_{3,1}$ DS.

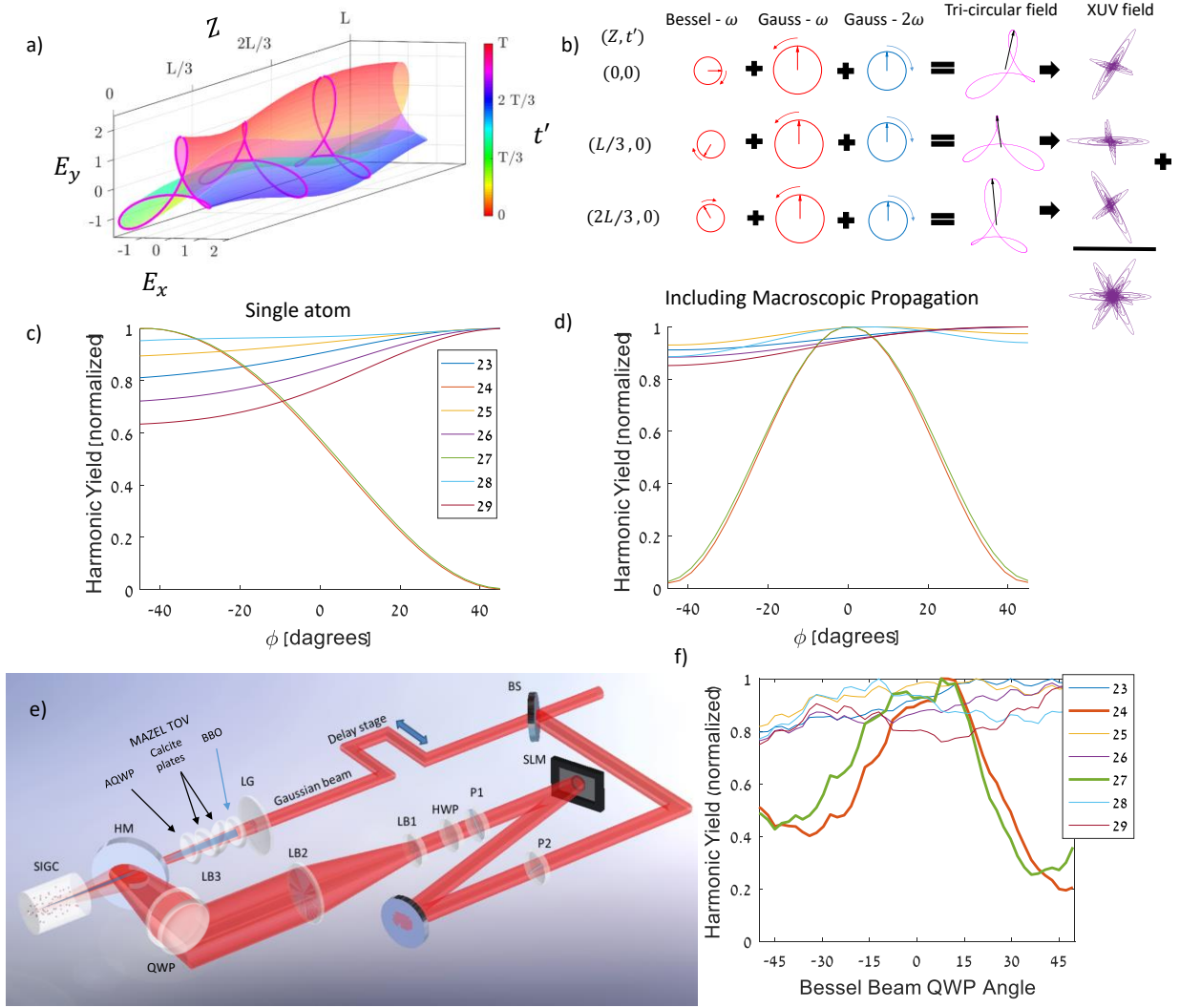

**Fig. 3.** a) The driver field (eq. 17) with $\varphi = -\pi/4$ plotted as a function of the propagation axis, Z, and retarded time, $t'$ (shown in color). Plotting the fields during the three propagation steps shows a three-fold $\hat{Z}_{3,1}\hat{\tau}_{3,1}\hat{r}_{3,-1}$ symmetry. b) The output XUV field is a coherent sum of all the emitted XUV fields during propagation and has three-fold $\hat{\tau}_{3,1}\hat{r}_{3,-1}$ symmetry, which results in the $3Q \pm 1$ selection rule. c) Normalized intensity of several harmonic orders as a function of $\phi$, calculated for only one propagation point, $\vec{E}_\varphi(t, z=0, \rho)$. The $3Q$ harmonics are forbidden only when $\varphi = 45°$. d) Normalized intensity of several harmonic orders as a function of $\varphi$, calculated for $\vec{E}_\varphi(t, z, \rho)$. The $3Q$ harmonics (e.g. harmonics 24,27) are forbidden at both $\varphi = -45°$ and $\varphi = 45°$, which correspond to the $\hat{e}_-$ and $\hat{e}_+$ circular polarizations of the Bessel beam, respectively. The $3Q$ harmonics are forbidden at $\varphi = 45°$ by the microscopic DS, $\hat{\tau}_{3,1}\hat{r}_{3,-1}$, and at $\varphi = -45°$ by the multi-scale DS, $\hat{Z}_{3,1}\hat{\tau}_{3,1}\hat{r}_{3,-1}$. e) Schematic plot of the experimental setup. The driver field is prepared by focusing the bi-circular $\omega - 2\omega$ Gaussian beams (produced by the MAZEL TOV apparatus) and the $\omega$ Bessel beam (shaped by the SLM) with controlled polarization (by the QWP) into the semi-infinite gas cell. f) Normalized intensity of several harmonic orders as a function of the Bessel QWP angle. Suppression of the $3Q$ harmonics (24,27) for both -45 and 45 degrees of the QWP, which leads to the $\hat{e}_-$ and $\hat{e}_+$ circular polarizations of the Bessel beam, respectively, are clearly observed.

Our experimental setup is illustrated in Fig. 3e. The output beam of a 1 kHz, 35 fs FWHM, 800nm carrier wavelength, Ti:sapphire amplifier (Coherent Legend USX) is split into two paths. The first beam, with 2.25 mJ per pulse, retains its spatial Gaussian profile and is focused with a lens ($f_{LG}$ = 300mm, Rayleigh range of the Gaussian beams = 5 mm) through the MAZEL TOV apparatus [38]. This apparatus consists of: (i) a

SHG crystal (0.5 mm thick BBO crystal) which transfers ~15% of the energy to the second harmonic beam, (ii) Calcite plates which pre-compensate for group delays induced by normally dispersive optics down the beam path, and (iii) a single achromatic quarter-wave waveplate (for the two spectral components) that converts the linear s-polarized fundamental and perpendicular p-polarized SH incoming beams to counter-rotating circularly polarized beams. This intense beam also drills a hole in the aluminum foil, terminating a semi-infinite gas cell (SIGC) filled with argon (45 torr). The second beam, with 0.5 mJ per pulse, undergoes an amplitude modulation using two perpendicularly oriented polarizers on either side of a phase-only spatial light modulator (SLM) (HoloeyePluto). This beam acquires the spatial distribution of a ring, which is then imaged and focused ($f_{LB3}$ = 150mm) to form a Bessel beam with $\beta/k = 0.0015$. Here, $k = 2\pi/800nm$ is the wave vector, and $\beta$ is the on-axis difference between the wave vectors of the Gauss and Bessel beams. This experimental condition corresponds to a periodicity of $L \approx 0.5mm$, which is one order of magnitude smaller than the Rayleigh range. A QWP is used to scan the polarization of the Bessel beam. Both beams are combined using a holed mirror and are focused close to the output of the SIGC where the high harmonic is generated. An aluminum filter downstream of the SIGS removes the pump beams before the spectrum of the HHG beam is measured by the XUV spectrometer. We first inserted only the bi-circular Gaussian beam fields (i.e., without the Bessel beam) and phased matched the HHG process by tuning the gas pressure, adjusting the location of the focus, and changing the opening of the aperture before the lens, to maximize the $3n \pm 1$ harmonics. Then, we added the Bessel beam and measured the harmonic intensity generated by the total driver field as a function of the QWP angle (i.e., $\varphi$). Fig. 3f shows the measured yield of harmonics 23-29 as a function of the QWP angle. (Each harmonic is normalized by its own peak intensity.) The suppression of the 3Q harmonic orders for both the QWP angles at $+45°$ and $-45°$ (yielding $\varphi = \pi/4$ and $\varphi = -\pi/4$ in Eq. (18), respectively) are clearly observed, as predicted from the selection rules by microscopic ($+45°$) and multi-scale ($-45°$) DSs.

**Discussion**

We presented a theory for symmetries and selection rules in (extreme) nonlinear optics for multi-scale systems. We introduced symmetries that couple time, macroscopic, and microscopic operations. We showed how these symmetries are transferred to the induced polarization and lead to constraints, i.e., selection rules, on physical observables. Multi-scale DSs and selection rules investigations in three different systems are presented: with spin-orbit nonlinear interaction, with quasi-periodic structures and experimental example with multi-scale DS in time, polarization and propagation axis.

A potential application of our theory is ultrafast spectroscopy the detection of the symmetry of the medium. This is done by using a driving field that has a DS. When the medium lacks that DS, the symmetry of the total system is reduced so that some restrictions in the selection rules are removed [32]. One important medium that breaks symmetries is a chiral medium, i.e., a medium that is asymmetric under any reflection or inversion. Probing the chirality of molecules can be a challenging task, and using multi-scale DS consideration can help in choosing the right microscopic [7,8] and macroscopic field parameters to enhance the far field chiral signal for molecular chirality detection and discrimination as shown in part 6 of the SI. Our work also paves the way for several interesting directions beyond harmonic generation. Extending the theory to non-local interactions may lead to new insights regarding multi-scale matter, light, and their interaction. Extensions to complexed structured laser ablation, to easily shaping complex structures should also be possible and exciting, leading to inducing symmetries in the media [39,40]. Overall, we expect that the use of multi-scale symmetries will lead to extended understanding of, and novel discoveries in, various multi-scale systems.

# Supplementary information: Multi-scale dynamical symmetries of electromagnetic fields and selection rules in nonlinear optics

Gavriel Lerner[1,2], Ofer Neufeld[1,2], Liran Hareli[3], Georgiy Shoulga[3], Eliayu Bordo[1,2], Avner Fleischer[4], Daniel Podolsky[1], Alon Bahabad[3] and Oren Cohen[1,2]

[1]*Physics Department, Technion – Israel Institute of Technology, Haifa, Israel.*
[2]*Solid State Institute, Technion – Israel Institute of Technology, Haifa, Israel.*
[3]*Department of Physical Electronics, Tel Aviv University, Tel Aviv, Israel.*
[4]*Chemistry Department, Tel Aviv University, Tel Aviv, Israel.*
*Corresponding authors' e-mail addresses: gavriel@campus.technion.ac.il, oren@si.technion.ac.il*

The supplementary information file contains the following sections: in supplementary note 1 we show how the DSs of the induced polarization arise from DSs of driver field. Supplementary note 2 shows how the selection rules in the induced polarization appear in the far-field HHG emission. Supplementary note 3 proves the connection between the DSs in real space to constrains in Fourier space. Supplementary note 4 derives the selection rules for macrospace and/or time inversion and reflection symmetry cases. Supplementary note 5 lists the macroscopic, time and microscopic building blocks operations that are used in the main text. Supplementary note 6 presents an application of our theory to derive selection rules for chiral dichroism in the interaction of locally-chiral light [1] with chiral media.

### 1) DSs of the induced polarization arising from DSs of the driver field

We will show here how, and in under which physical conditions, the multi-scale DSs of the driver field are transferred to the induced polarization. The dynamics of an electronic (with N electrons) wave function of a system with the Hamiltonian in eq. (4) of the main text, $\psi_{\vec{R}}(\vec{r}_1, \vec{r}_2, ..., \vec{r}_N, t)$, is governed by the following TDSE:

$$i\frac{\partial}{\partial t}\psi_{\vec{R}}(\vec{r}_1, ..., \vec{r}_N, t) = \left[\sum_j \frac{\nabla_j^2}{2} + \frac{1}{2}\sum_{i \neq j}|\vec{r}_i - \vec{r}_j|^{-1} + \sum_j U_{\vec{R}}(\vec{r}_j) + \sum_j \vec{E}(\vec{R}, t) \cdot \vec{r}_j\right]\psi_{\vec{R}}(\vec{r}_1, ..., \vec{r}_N, t) \quad (1)$$

Applying the coordinate transformation, $\vec{r}_j \to \hat{\gamma}\vec{r}_j$ (i.e., a point group operation, $\hat{\gamma}$, operating on the microscopic space), and time transformation, $t \to st + \tau$, where $\tau$ is the time translation and $s = \pm 1$ (-1 for time-reversal), on eq. (1) yields

$$i\frac{\partial}{\partial t}\psi_{\vec{R}}(\hat{\gamma}\vec{r}_1, ..., \hat{\gamma}\vec{r}_N, st + \tau)$$
$$= \left[\sum_j \frac{\nabla_j^2}{2} + \frac{1}{2}\sum_{i \neq j}|\hat{\gamma}\vec{r}_i - \hat{\gamma}\vec{r}_j|^{-1} + \sum_j U_{\vec{R}}(\hat{\gamma}\vec{r}_j) + \sum_j \vec{E}(\vec{R}, st + \tau) \cdot \hat{\gamma}\vec{r}_j\right]\psi_{\vec{R}}(\hat{\gamma}\vec{r}_1, ..., \hat{\gamma}\vec{r}_N, st + \tau) \quad (2)$$

If the electric field is conserved under the multi-scale DS (i.e., $\hat{\gamma}\vec{E}(\vec{R}, st + \tau) = \vec{E}(\widehat{M}_R\vec{R}, t)$) and the microscopic potential of the medium shares the same symmetry (i.e., the relation between the microscopic static potential in points $\vec{R}$ and $\widehat{M}_R\vec{R}$ is $U_{\vec{R}}(\hat{\gamma}\vec{r}) = U_{\widehat{M}_R\vec{R}}(\vec{r})$, as is the case with homogeneous and isotropic media), then:

$$i\frac{\partial}{\partial t}\psi_{\vec{R}}(\hat{\gamma}\vec{r}_1, ..., \hat{\gamma}\vec{r}_N, st + \tau)$$
$$= \left[\sum_j \frac{\nabla_j^2}{2} + \frac{1}{2}\sum_{i \neq j}|\vec{r}_i - \vec{r}_j|^{-1} + \sum_j U_{\widehat{M}_R\vec{R}}(\vec{r}_j) + \sum_j \vec{E}(\widehat{M}_R\vec{R}, t) \cdot \vec{r}_j\right]\psi_{\vec{R}}(\hat{\gamma}\vec{r}_1, ..., \hat{\gamma}\vec{r}_N, st + \tau) \quad (3)$$

such that

$$i\frac{\partial}{\partial t}\widehat{G}'\psi_{\vec{R}}(t) = \widehat{H}_{\widehat{M}_R\vec{R}}(t)\widehat{G}'\psi_{\vec{R}}(t) \tag{4}$$

where $\widehat{G}'$ is a unitary (or antiunitary, in the case of time reversal) transformation of time and microspace, where $\widehat{G}'\,\widehat{M}_R = \widehat{G}$. Now, if we look at the full Hamiltonian of the non-interacting microscopic systems, $\widehat{H}(t) = \sum_{\vec{R}}\widehat{H}_{\vec{R}}(t)$, which describes the dynamics of the full wave function of non-interacting microscopic systems, $\psi(t) = \prod_{\vec{R}}\psi_{\vec{R}}(t)$, we see that according to eq. (4), the full Hamiltonian is conserved under the multi-scale DS:

$$\widehat{G}\widehat{H}(t)\widehat{G}^{-1} = \sum_{\vec{R}}\widehat{G}'\,\widehat{H}_{\widehat{M}_R\vec{R}}(t)\,\widehat{G}'^{-1} = \sum_{\vec{R}}\widehat{H}_{\vec{R}}(t) = \widehat{H}(t) \tag{5}$$

Therefore we can use Floquet theory in a similar manner as in [2]. For a periodic Hamiltonian (if the EM field is aperiodic with multiple frequencies, then the system can by described by the many-mode Floquet state theory [3,4]), $\widehat{H}(t+T) = \widehat{H}(t)$, and the Floquet Hamiltonian, $\mathcal{H}_F = \widehat{H}(t) - i\frac{\partial}{\partial t}$, has eigenstates which are T-periodic Floquet modes, $u_n(t)$, with corresponding quasi-energies, $\varepsilon_n$. Solutions to the time-dependent Schrödinger equation (TDSE) of the full Hamiltonian, $\widehat{H}(t)$, are comprised of Floquet states $\psi_n(t) = e^{i\varepsilon_n t}u_n(t)$. If $\widehat{G}\widehat{H}(t) = \widehat{H}(t)$ (eq. (5)), then $[\widehat{G}, \mathcal{H}_F] = 0$, such that the Floquet modes are simultaneous eigenmodes of the Floquet Hamiltonian and $\widehat{G}$. Furthermore, since $\widehat{G}$ is unitary or anti-unitary, its eigenvalues are roots of unity; $\widehat{G}u_n(t) = e^{i\phi_n}u_n(t)$, where $\phi_n$ is real. Hence, if the initial wave function populates a single Floquet state, any measured observable (e.g. the induced polarization), $o(\vec{R},t) = \langle u(t)|\widehat{O}|u(t)\rangle = \langle u(t)|\sum_{\vec{R}}\widehat{O}_{\vec{R}}|u(t)\rangle$, also upholds the DS:

$$o(\vec{R},t) = \langle u(t)|\widehat{G}^{-1}\widehat{G}\widehat{O}\widehat{G}^{-1}\widehat{G}|u(t)\rangle = \widehat{G}o(\vec{R},t) \tag{6}$$

The EM field is often a pulse with finite duration, so the requirement becomes that the wave function before the pulse exhibits the DS, which is the case for an isotropic gas in the ground state. The pulse should be turned on adiabatically to approximately initiate steady-state dynamics of a single Floquet state [5]. Another requirement is that the temporal frequencies of the driver and HHG emission are not close to a resonance frequency of the potential, $U$ [5]. Also, if the initial state is not a single Floquet mode, but a superposition of Floquet modes, eq. (6) will not hold unless the medium consists of an isotropic ensemble that contains an equal population of all degenerate states [2]. Under the above conditions, the symmetry of the field, $\vec{E}(\widehat{M}_R\vec{R},t) = \widehat{G}'\vec{E}(\vec{R},t)$, is transferred to the induced polarization, $\vec{P}(\widehat{M}_R\vec{R},t) = \widehat{G}'\vec{P}(\vec{R},t)$.

Notably, the above discussion did not include macroscopic effects that could also affect the symmetry-relations and possible break the symmetry (e.g. due to phase-matching, spatial averaging, etc.). They are discussed in the next section.

### 2) Selection rules in the induced polarization and their appearance in the far field of the HHG

In the main text, we discussed how DSs lead to selection rules in the Fourier components of the induced polarization. Here we discuss and analyze how this selection will appear in the far field of the harmonic generation, which is measured in HHG experiments.

We first discuss the macroscopic conditions under which the DSs of the driver field are transferred to the induced polarization. The principle condition is that the entire system (driver field and medium) should be conserved under the multi-scale DSs. Imperfections in these symmetries (e.g., due to non-uniform density of the gas) lead to deviations from the symmetry and selection rules. In a case where the symmetry includes propagation, in order that the driver field will be periodic in the propagation axis, it needs to be approximately non-depleted and loosely focused (such that the Rayleigh length of the focusing beam is larger than the interaction length). We point out that the driver field (i.e., $\vec{E}(\vec{R},t)$) is the field inside the interaction region which is sensitive to the medium dispersion. Moreover, phase mismatch between the induced polarization and emitted harmonic field can alter the transfer of the selection rules to the far field.

Now we will analyze how the selection rules in the Fourier components of the induced polarization will appear in the far field of the harmonic generation. The induced polarization radiates the HHG field, $\vec{E}_{HHG}$, according to the inhomogeneous wave equation:

$$\left(\nabla^2 - \frac{1}{c^2}\partial_t^2\right)\vec{E}_{HHG}(X,Y,Z,t) = \frac{4\pi}{c^2}\partial_t^2\vec{P}(X,Y,Z,t) \tag{7}$$

This equation can be solved numerically using the discrete dipole approximation [6]. The detection of the harmonics is typically performed at approximately one-meter scale after the interaction region, and the diameter of the detector is several centimeters. (In many HHG setups there is also a small aperture between the interaction region and the detector which also restricts the detection to small divergence.) Therefore, the detection captures the far-field of the

HHG emission which can be approximated by the Fraunhofer diffraction equation for each Z slice of the interaction region with a total length, $L$:

$$\vec{E}_{HHG}(x_d, y_d, \omega) \simeq \int_0^L dZ \frac{k^2 e^{ik(z_d-Z)} e^{\frac{ik(x_d^2+y_d^2)}{2(z_d-Z)}}}{2\pi i(z_d - Z)} \iint dXdY e^{\frac{-ik}{z_d-Z}(x_d X + y_d Y)} \vec{P}(X,Y,Z,\omega) \tag{8}$$

where $x_d$ and $y_d$ are the coordinates of the detector that is located $z_d$ distance from the interaction region, and $k$ is the wavenumber of temporal frequency, $\omega$. (Outside the interaction region $k = \omega/c$, and inside the interaction region $k = n_\omega \omega/c$ ($|n_\omega - 1| \ll 1$, is typically the case in the XUV.) By using the condition $\frac{Lkx_d X}{z_d^2} \lesssim 0.01 q_\omega \ll 1$, we can approximate:

$$\vec{E}_{HHG}(x_d, y_d, \omega) \approx \frac{\omega^2 e^{i\frac{\omega}{c} z_d \left(1 + \frac{x_d^2 + y_d^2}{2z_d^2}\right)}}{2\pi i z_d c^2} \int_0^L dZ e^{-i\frac{n_\omega \omega}{c} Z \left(1 - \frac{x_d^2 + y_d^2}{2z_d^2}\right)} \iint dXdY e^{\frac{-in_\omega \omega}{cz_d}(x_d X + y_d Y)} \vec{P}(X,Y,Z,\omega) \tag{9}$$

The last term is the 2D far field Fourier transform with frequencies $kx_d/2\pi z_d$ and $ky_d/2\pi z_d$. This equation is easy to calculate numerically for a given $\vec{P}(X,Y,Z,\omega)$. However, the simulation of $\vec{P}(X,Y,Z,\omega)$ is difficult since it involves solving the TDSE for many $(X,Y,Z)$ points. However, by using the DS, it is enough to calculate $\vec{P}(X,Y,Z,\omega)$ only inside the spatial unit cell of the DS and then extrapolate $\vec{P}(X,Y,Z,\omega)$ outside the unit cell according to the DS. Here, we express the far field emission using the Fourier series of $\vec{P}(X,Y,Z,t)$ with a (3+1)D Gaussian envelope:

$$\vec{P}(X,Y,Z,t) = e^{-\frac{X^2}{2\sigma_X^2} - \frac{Y^2}{2\sigma_Y^2} - \frac{t^2}{2\sigma_t^2}} \sum_{\vec{k}=(k_x,k_y,k_z,\omega_q)} \vec{F}(\vec{k}) \exp\left(i(k_x X + k_y Y + k_z Z - \omega_q t)\right) \tag{10}$$

which plugged into eq. (9), gives:

$$\vec{E}_{HHG}(x_d, y_d, \omega) \approx \sum_{\vec{k}} \left\{ \vec{F}(\vec{k}) \frac{\omega^3 \sigma_X \sigma_Y \sigma_t L e^{i\frac{\omega}{c} z_d \left(1 + \frac{x_d^2 + y_d^2}{2z_d^2}\right)}}{2\pi i z_d c^2} g_\omega(\omega - \omega_q) \text{sinc}\left(L\left(k_z - \frac{n_\omega \omega}{c}\left(1 - \frac{x_d^2 + y_d^2}{2z_d^2}\right)\right)\right) g_{k_x}\left(k_x - \frac{x_d}{z_d}\frac{n_\omega \omega}{c}\right) g_{k_y}\left(k_y - \frac{y_d}{z_d}\frac{n_\omega \omega}{c}\right) \right\} \tag{11}$$

where $g_\omega(\omega) = e^{-\frac{\sigma_t^2 \omega^2}{2}}$, $g_{k_x}(k_x) = e^{-\frac{\sigma_X^2 k_x^2}{2}}$, and $g_{k_y}(k_y) = e^{-\frac{\sigma_Y^2 k_y^2}{2}}$ and $\sigma_X$, $\sigma_Y$, and $\sigma_t$ are the spatial and temporal Gaussian widths, respectively. This expression is maximal for $k_z = \frac{n_\omega \omega}{c}\left(1 - \frac{x_d^2 + y_d^2}{2z_d^2}\right), k_x = \frac{n_\omega \omega}{c}\frac{x_d}{z_d}, k_y = \frac{n_\omega \omega}{c}\frac{y_d}{z_d}$, and $\omega = \omega_q$ which gives the condition of phase matching, $k_z^2 + k_x^2 + k_y^2 = \left(\frac{n_{\omega_q} \omega_q}{c}\right)^2$.

By considering this phase matching condition, in the case of our experiment with the tri-circular beam (as described in Fig. 3 of the main text), we can see that $k_z = \frac{q_1 \omega}{c} - q_2 \beta$ of the induced polarization is phased matched for $q_2 \beta = \frac{q_1 \omega}{c} \frac{x_d^2 + y_d^2}{2z_d^2}$. Therefore, the on-axis harmonics are phase matched only for $q_2 = 0$. (Hence, the selection rule for the temporal harmonics is $q_1 = 3n \pm 1$, as was demonstrated in the experiment.) Rings with radius $\sqrt{\frac{2z_d^2 q_2 \beta c}{q_1 \omega}}$ are also allowed (with temporal harmonic, $q_1 = 3n \pm 1 - q_2$, as was mentioned in the main text).

Another consideration is reabsorption of the emitted high harmonic field. The index of refraction in the XUV is the complex valued $n_{\omega_q} = 1 + \Delta n_{\omega_q} + i\beta_{\omega_q}$, with typical values (far from resonance) $|\Delta n_{\omega_q}| \lesssim 10^{-5}$ [7] and $\beta_{\omega_q} \lesssim 10^{-6}$ (or $k' = \beta_{\omega_q} \frac{\omega}{c} \lesssim 1 mm^{-1}$) [8] for 1 bar pressure. Inserting the complex valued $n_{\omega_q}$ into the $Z$-dependent part in eq. (9) gives:

$$\int_0^L dZ e^{(i\Delta k - \tilde{k})Z} = \frac{\tilde{k} + e^{-\tilde{k}L}\left(\Delta k \sin(\Delta kL) - \tilde{k}\cos(\Delta kL)\right)}{\tilde{k}^2 + \Delta k^2} \tag{12}$$

where $\Delta k = k_z - \frac{(1+\Delta n_{\omega q})\omega}{c}\left(1 - \frac{x_d^2 + y_d^2}{2z_d^2}\right)$ and $\tilde{k} = \frac{\beta_{\omega q}\omega}{c}\left(1 - \frac{x_d^2 + y_d^2}{2z_d^2}\right)$. Fig. 1 shows the amplitude of an emitted harmonic as a function of $\tilde{k}L$ and $\Delta kL$ according to eq. (12). When the absorption is small, i.e., $\tilde{k} < \Delta k, 1/L$, eq. (12) reduces to $L\mathrm{sinc}(\Delta kL)$. When the absorption is large, i.e., $\tilde{k} > 1/L$, eq. (12) reduces to $\frac{\tilde{k}}{\tilde{k}^2 + \Delta k^2}$. This means that as $\tilde{k}L$ gets larger, the phase matching condition for $\Delta kL$ is less strict. Therefore, when the absorption is large, harmonics with larger values of $\Delta k$ can have significant amplitudes compared to harmonics with small values of $\Delta k$. In the case of our experimental example, this also means that harmonics with $q_2 = 1$ can arrive to the detector with temporal harmonics $q_1 = 3n \pm 1 - q_2$. In this case, the amplitude of the $q_1 = 3n$ harmonics will not drop completely to zero, as can be seen in fig. 3f in the main text.

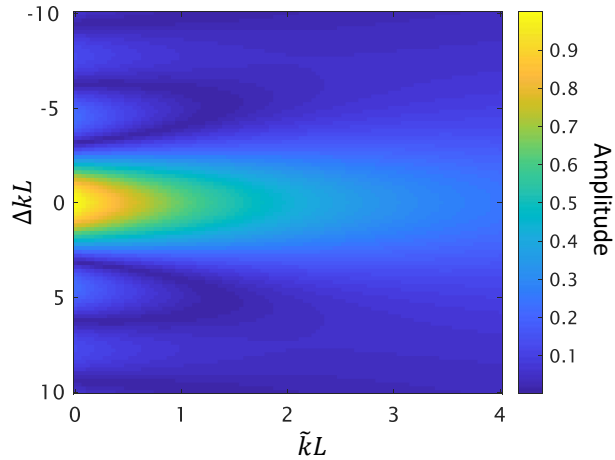

**Fig. 1.** The joint effect of phase mismatch $\Delta k$ and absorption $\tilde{k}$ on the amplitude, according to eq. (12).

3) **Constraints on the Fourier spectrum of the induced polarization from DSs in real space**

We will now prove that a general DS $\hat{G} = \hat{\gamma}\hat{M}$ leads to eq. (6) in the main text which gives constraints on the harmonics of the induced polarization $\vec{P}$ in the Fourier space.

$$\vec{P}(\vec{X}) = \hat{G}^{-1}\vec{P}(\vec{X}) = \hat{\gamma}^{-1}\vec{P}(\widehat{M}\vec{X}) = \hat{\gamma}^{-1}\vec{P}(\hat{\Gamma}^{-1}\vec{X} - \vec{a}) \tag{13}$$

In the Fourier domain, this constraint reads:

$$\sum_{\vec{k}} \vec{F}(\vec{k})\exp(i\vec{k}\cdot\vec{X}) = \sum_{\vec{k}} \hat{\gamma}^{-1}\vec{F}(\vec{k})\exp(i\vec{k}\cdot(\hat{\Gamma}^{-1}\vec{X} - \vec{a})) \tag{14}$$

The left hand side of eq. (14) can also be written, by a change of the order of summation, as

$$\begin{aligned}\sum_{\vec{k}} \vec{F}(\hat{\Gamma}\vec{k})\exp(i\hat{\Gamma}\vec{k}\cdot\vec{X}) &= \sum_{\vec{k}} \vec{F}(\hat{\Gamma}\vec{k})\exp(i\hat{\Gamma}^{-1}\hat{\Gamma}\vec{k}\cdot\hat{\Gamma}^{-1}\vec{X}) \\ &= \sum_{\vec{k}} \vec{F}(\hat{\Gamma}\vec{k})\exp(i\vec{k}\cdot\hat{\Gamma}^{-1}\vec{X}) = \sum_{\vec{k}}\hat{\gamma}^{-1}\vec{F}(\vec{k})\exp(i\vec{k}\cdot(\hat{\Gamma}^{-1}\vec{X} - \vec{a}))\end{aligned} \tag{15}$$

Consequently, the Fourier coefficients of identical exponents $\exp(i\vec{k}\cdot\hat{\Gamma}^{-1}\vec{X})$ in Eq. (15) must be equal and therefore $\vec{F}(\hat{\Gamma}\vec{k}) = \hat{\gamma}^{-1}\vec{F}(\vec{k})\exp(-i\vec{k}\cdot\vec{a})$, and after multiplication with $\hat{\gamma}\exp(i\vec{k}\cdot\vec{a})$:

$$\hat{\gamma}\vec{F}(\hat{\Gamma}\vec{k})\exp(i\vec{k}\cdot\vec{a}) = \vec{F}(\vec{k}) \tag{16}$$

Which is Eq. (7) in the main text.

4) **Selection rules of macrospace inversion and reflection and/or time reversal**

We now derive the selection rules for symmetries that include time reversal or space reflection (rows 9-14 in table 1 in the main text for the case of (1+1+2)D).

**DS with macrospace-inversion and time reversal**

Since macrospace-inversion and time reversal (i.e. $\hat{\Gamma}\vec{X} = -\vec{X}$) is an order two operator, then the microscopic operation $\hat{\gamma}$ must be of order one or two, which can be reflection, inversion or 180 degrees rotations. Therefore, the eigenvector polarizations of $\hat{\gamma}$, $\hat{F}_i(\vec{k})$, are orthogonal linear polarizations: $\hat{F}_x(\vec{k})$, $\hat{F}_y(\vec{k})$, and $\hat{F}_z(\vec{k})$ (where with no loss of generality the x,y,z coordinates are chosen such that one of them is along the rotation axis or normal to the reflection plane of $\hat{\gamma}$) with possible $\alpha_i$ eigenvalues of 0 or $\pm\pi$. Since $\vec{P}(\vec{X})$ is real, $\vec{F}(-\vec{k})$ equals the complex conjugate $\vec{F}^*(\vec{k})$; therefore, $\phi_i(-\vec{k}) = -\phi_i(\vec{k})$. The phase difference between two polarizations with the same $\vec{k}$ will be:

$$\Delta\phi_{i,i'} = \phi_i(\vec{k}) - \phi_{i'}(\vec{k}) = \vec{k}\cdot\vec{a} + \alpha_i - \phi_i(\vec{k}) - \left(\vec{k}\cdot\vec{a} + \alpha_{i'} - \phi_{i'}(\vec{k})\right) = \alpha_i - \alpha_{i'} - \Delta\phi_{i,i'} \qquad (17)$$

hence:

$$\Delta\phi_{i,i'} = (\alpha_i - \alpha_{i'})/2 \qquad (18)$$

According to eq. (18), the phase difference between the two linear polarization components (i.e., $\Delta\phi_{i,i'}$) of the $\vec{k}$ harmonics equals $(\alpha_i - \alpha_{i'})/2$, which is expressed as either (i) a multiple of $\pi$, which means that the two linear polarization components, $\hat{F}_i(\vec{k})$ and $\hat{F}_{i'}(\vec{k})$, are in phase, and thus polarization is linear in the $i - i'$ plane; or (ii) $\pm\pi/2$, in which case all the harmonics are elliptically polarized with minor/major ellipse axes along $i$ and $i'$. Therefore, for N=2, if $\hat{\gamma}$ is the identity or $\hat{r}_2$ operation, then all the harmonics are linearly polarized (row 9 table 1 in the main text). Otherwise, if $\hat{\gamma}$ is a reflection operator, then all of the harmonics are elliptically polarized with major/minor axes corresponding to the reflection axes (row 10 in table 1 in the main text). For N=3, if $\hat{\gamma}$ is the identity or inversion operation, all harmonics are linearly polarized. Instead, if $\hat{\gamma} = \hat{r}_2$, the rotation axis is a major/minor axis of the polarization ellipsoid. If $\hat{\gamma}$ is a reflection operator, the polarization ellipsoid has a major/minor axis normal to the reflection plane.

**DS with macrospace reflection**

Now we will consider DSs that do not involve macrospace-inversion and time reversal, but require only some reflection in macrospace or time. If the DS involves a macrospace reflection, then eq. (8) in the main text becomes:

$$A_{q_1,-q_2}\exp(i\phi_{q_1,-q_2} + i\vec{k}\cdot\vec{a})\hat{\gamma}\hat{F}_{q_1,-q_2} = A_{q_1,q_2}\exp(i\phi_{q_1,q_2})\hat{F}_{q_1,q_2} \qquad (19)$$

Therefore, the amplitudes of the mirrored spatial harmonics, $q_2$ and $-q_2$, are the same, i.e., $A_{q_1,q_2} = A_{q_1,-q_2}$. According to eq. (19), when the microscopic operation ($\hat{\gamma}$) is the identity or the inversion operations ($\pi$ rotation in 2D), $\hat{F}_{q_1,-q_2}$ must equal $\hat{F}_{q_1,q_2}$ (row 11 in table 1 in the main text). Also, if $\hat{\gamma}$ is a reflection operation, $\hat{\sigma}$, then $\hat{\sigma}\hat{F}_{q_1,-q_2} = \hat{F}_{q_1,q_2}$ (row 12 in table 1 in the main text). Eq. (19) also determines the relationship between the phases $\phi_{q_1,-q_2}$ and $\phi_{q_1,q_2}$. This relationship depends on $\hat{\gamma}$ and the propagation term, $\vec{k}\cdot\vec{a}$.

**DS with time reversal**

When the DS involves a time reversal, eq. (8) in the main text becomes:

$$A_{-q_1,q_2}\exp(i\phi_{q_1,-q_2} + i\vec{k}\cdot\vec{a})\hat{\gamma}\hat{F}_{q_1,-q_2} = A_{q_1,q_2}\exp(i\phi_{q_1,q_2})\hat{F}_{q_1,q_2} \qquad (20)$$

Therefore, $A_{q_1,q_2} = A_{-q_1,q_2}$. When $\hat{\gamma}$ is the identity or the inversion operation ($\pi$ rotation in 2D), $\hat{F}_{-q_1,q_2} = \hat{F}_{q_1,q_2}$. Hence, the polarization must be linear (row 13 in table 1 in the main text). When $\hat{\gamma}$ is the reflection operation, $\hat{\sigma}$, then $\hat{\sigma}\hat{F}_{-q_1,q_2} = \hat{F}_{q_1,q_2}$, and the polarization ellipse axis must be orthogonal to the $\hat{\sigma}$ plane (row 14 in table 1 in the main text).

5) **Tables of macroscopic, time and microscopic building blocks operations**

In the main text we described multi-scale DS and discussed various examples. All symmetries for the (2+1+1)d case were outlined in table 1 of the main text. The number of all possible symmetries (and the number of symmetry groups) grows exponentially as the dimensionality increases. The enumeration and classification of all the symmetries and symmetry groups can be done in a similar manner as was done for magnetic groups [9]. The building blocks operations for multi-scale DS are describe in the main text, and listed here in tow tables: table 1 for the macrospace-time operation and table 2 for microspace operations. In table 2 we also give the eigenvectors and eigenvalues which are used in the selection rules derivations.

**Table 1.** Macrospace and time operation building blocks, their actions on $\vec{E}(\vec{R},t)$, and their associated order.

| Macroscopic-time operation $\hat{M}$ | $\hat{M}\vec{E}(\vec{R},t)$ | Order of $\hat{M}$ |
|---|---|---|
| Time translation, $\hat{t}_{n,m}$ | $\vec{E}(\vec{R}, t + Tm/n)$ | $n$ |
| Space translation, $\hat{J}_{n,m}$ | $\vec{E}(\vec{R} + Lm/n\,\hat{j}, t)$ | $n$ |
| Macrospace-time translation, $\hat{D}_n$ | $\vec{E}\left(\vec{R} + \dfrac{\vec{u}}{n}, t + \tau/n\right)$ | $n$ |
| Time-reversal, $\hat{T}$ | $\vec{E}(\vec{R}, -t)$ | 2 |
| Reflection, $\hat{\Sigma}_x$ | $\vec{E}(-X, Y, Z, t)$ | 2 |
| Space rotation, $\hat{R}_{n,m}$ | $\vec{E}(X', Y', Z, t)$ | $n$ |

**Table 2.** Microscopic operation building blocks, e.g., reflection and rotation. The order, matrix representation, eigenvectors, and eigenvalues of each operation is listed. The rotation angle is defined as $\theta = 2\pi m/n$.

| Microscopic Operation, $\hat{\gamma}$ | Order | Matrix Representation | Eigenvectors, $\hat{F}_{\vec{n}}^{(m)}$ | Eigenvalues, $e^{i\alpha^{(m)}}$ |
|---|---|---|---|---|
| Reflection, $\hat{\sigma}_h$ | 2 | $\begin{pmatrix} -1 & 0 & 0 \\ 0 & 1 & 0 \\ 0 & 0 & 1 \end{pmatrix}$ | $\begin{pmatrix}1\\0\\0\end{pmatrix}, \begin{pmatrix}0\\1\\0\end{pmatrix}, \begin{pmatrix}0\\0\\1\end{pmatrix}$ | $\pi, 0, 0$ |
| Rotation, $\hat{r}_{n,m}$ | $n$ | $\begin{pmatrix} \cos(\theta) & -\sin(\theta) & 0 \\ \sin(\theta) & \cos(\theta) & 0 \\ 0 & 0 & 1 \end{pmatrix}$ | $\begin{pmatrix}1\\+i\\0\end{pmatrix}, \begin{pmatrix}1\\-i\\0\end{pmatrix}, \begin{pmatrix}0\\0\\1\end{pmatrix}$ | $+\theta, -\theta, 0$ |
| Elliptical rotation, $\hat{e}_{n,m}$ | $n$ | $\begin{pmatrix} \cos(\theta) & -\sin(\theta)/\epsilon & 0 \\ \epsilon\sin(\theta) & \cos(\theta) & 0 \\ 0 & 0 & 1 \end{pmatrix}$ | $\begin{pmatrix}1\\+i\epsilon\\0\end{pmatrix}, \begin{pmatrix}1\\-i\epsilon\\0\end{pmatrix}, \begin{pmatrix}0\\0\\1\end{pmatrix}$ | $+\theta, -\theta, 0$ |
| Elliptical improper rotation, $\hat{s}_{n,m}$ | $n$ for even $n$, $2n$ for odd $n$ | $\begin{pmatrix} \cos(\theta) & -\sin(\theta)/\epsilon & 0 \\ \epsilon\sin(\theta) & \cos(\theta) & 0 \\ 0 & 0 & -1 \end{pmatrix}$ | $\begin{pmatrix}1\\+i\epsilon\\0\end{pmatrix}, \begin{pmatrix}1\\-i\epsilon\\0\end{pmatrix}, \begin{pmatrix}0\\0\\1\end{pmatrix}$ | $+\theta, -\theta, \pi$ |

### 6) Selection rules for chiral dichroism in HHG

This section presents an application of our multi-scale DS theory for analyzing enantio-sensitive HHG [10,11]. In a recent work [1], it was shown that HHG driven by "synthetic chiral light" – i.e. light with electric field vector that does not exhibit any microscopic DS involving reflection, inversion or improper rotations – display enantio-sensitive HHG power spectrum in the near field. When the synthetic chiral light is also "globally chiral", i.e. it has the same handedness in the entire interaction region, the far-field HHG is also enantio-sensitive. The situation is less clear when the driving field is synthetically chiral, but not globally chiral. Later work [12] demonstrated that some such fields do exhibit far-field enantio-sensitive HHG spectra. We show below that multi-scale DS theory can be used for analyzing these cases, providing chiral dichroism (CD) selection rules and physical insight.

We first analyzed the field presented in Ref. [12] which is given by (see eq. 21-27 in Ref. [12]):

$$\vec{E}(t,X,Y) = E_\omega\left[E_x(X)\hat{x} - iE_y(Y)\hat{y}\right]e^{i(k_yY-\omega t)} + E_{2\omega}E_z(X)\hat{z}e^{2i(k_yY-\omega t+\phi)} + c.c. \quad (21)$$

where

$$\begin{aligned} E_x(X) &= \cos(\alpha)\cos(k_xX) \\ E_y(X) &= \sin(\alpha)\sin(k_xX) \\ E_z(X) &= \cos(\alpha)\cos(2k_xX) \end{aligned} \quad (22)$$

where $\alpha$ is the half cross angle between the beams propagating in the x-y plane, $E_{m\omega}$ is the amplitude of the $m\omega$ ($m = 1,2$) field and $k_x = \sin(\alpha)\, k$. The field is locally-chiral (i.e. has a nonzero degree of chirality [13]) but not globally-chiral (i.e. the sign of the field's chirality changes rapidly across the interaction region). Through numerical simulations, Ref. [12] found that some harmonic orders still exhibit large chiral dichroism in the far field that is anti-symmetric with respect to the propagation angle (Fig. 4 in Ref. [12]). We show here that multi-scale DSs of the light-matter system lead to a selection rule of the chiral dichroism spectra that matches and explains this feature.

We have found that the field in Eq. (21) exhibits the following multi-scale symmetry: $\hat{\sigma}_{xz}\hat{\Sigma}$, i.e. $\hat{\sigma}_{xz}\vec{E}(t,-X) = \vec{E}(t,X)$. The microscopic reflection $\hat{\sigma}_{xz}$ flips the handedness of the field; hence, the interaction of $\vec{E}(t,X)$ with a medium of randomly oriented left-handed chiral molecules is equivalent to the interaction of $\vec{E}(t,-X)$ with randomly oriented right-handed chiral molecules up to a microscopic reflection $\hat{\sigma}_{xz}$. Therefore, the induced polarization of the two ennatiomers exhibits the relation:

$$\hat{\sigma}_{xz}\vec{P}^L(t,-X) = \vec{P}^R(t,X) \tag{23}$$

Implementing eq. (7) of the main text on eq. (23) leads to:

$$\hat{\sigma}_{xz}\vec{F}^L(q_1\omega,-q_2k_x) = \vec{F}^R(q_1\omega,q_2k_x) \tag{24}$$

The chiral dichroism of the far field HHG is defined as

$$CD(q_1\omega, q_2k_x) = 2\frac{I^L(q_1\omega,q_2k_x) - I^R(q_1\omega,q_2k_x)}{I^L(q_1\omega,q_2k_x) + I^R(q_1\omega,q_2k_x)} = 2\frac{\left|\vec{F}^L(q_1\omega,q_2k_x)\right|^2 - \left|\vec{F}^R(q_1\omega,q_2k_x)\right|^2}{\left|\vec{F}^L(q_1\omega,q_2k_x)\right|^2 + \left|\vec{F}^R(q_1\omega,q_2k_x)\right|^2} \tag{25}$$

where $I^L$ ($I^R$) is the intensity of emission from the left (right) handed enantiomer. Using the relation in Eq. (24) we obtain.

$$\begin{aligned}CD(q_1\omega,-q_2k_x) &= 2\frac{\left|\vec{F}^L(q_1\omega,-q_2k_x)\right|^2 - \left|\vec{F}^R(q_1\omega,-q_2k_x)\right|^2}{\left|\vec{F}^L(q_1\omega,-q_2k_x)\right|^2 + \left|\vec{F}^R(q_1\omega,-q_2k_x)\right|^2} \\ &= 2\frac{\left|\hat{\sigma}_{xz}\vec{F}^R(q_1\omega,q_2k_x)\right|^2 - \left|\hat{\sigma}_{xz}\vec{F}^L(q_1\omega,q_2k_x)\right|^2}{\left|\hat{\sigma}_{xz}\vec{F}^R(q_1\omega,q_2k_x)\right|^2 + \left|\hat{\sigma}_{xz}\vec{F}^L(q_1\omega,q_2k_x)\right|^2} \\ &= -2\frac{\left|\vec{F}^L(q_1\omega,q_2k_x)\right|^2 - \left|\vec{F}^R(q_1\omega,q_2k_x)\right|^2}{\left|\vec{F}^L(q_1\omega,q_2k_x)\right|^2 + \left|\vec{F}^R(q_1\omega,q_2k_x)\right|^2} = -CD(q_1\omega,q_2k_x)\end{aligned} \tag{26}$$

Eq. (26) establishes the following selection rule for CD: for a given emitted harmonic with temporal frequency $q_1\omega$, the CD at opposite spatial frequencies $q_2k_x$ and $-q_2k_x$ have opposite sign. This feature is clearly seen in Fig. 4 of Ref. [12].

Next, we analyzed another field that is locally-chiral but not globally-chiral, which was described in the supplementary of ref. [1] and analyzed in ref. [13]. The field is a superposition of two non-collinear counter-rotating bi-elliptical fields, which near the focus are given by (eq S34-S5 in SI of Ref. [1]):

$$\begin{aligned}\vec{E}_1(t,X,Y) &= \frac{1}{2}E_{1,0}\exp(ik(\sin(\alpha)X + \cos(\alpha)Y) - i\omega t)(\cos(\alpha)\hat{x} - \sin(\alpha)\hat{y} + i\epsilon_1\hat{z}) + c.c. \\ \vec{E}_2(t,X,Y) &= \frac{1}{2}E_{2,0}\exp(i2k(-\sin(\alpha)X + \cos(\alpha)Y) - i2\omega t)(\cos(\alpha)\hat{x} + \sin(\alpha)\hat{y} - i\epsilon_2\hat{z}) + c.c.\end{aligned} \tag{27}$$

Where $E_{n,0}$ is the electric amplitude $\epsilon_n$ is the ellipticity of the two fields.

Through a rigorous analysis, we have found that the total field $\vec{E} = \vec{E}_1 + \vec{E}_2$ exhibits the following multi-scale DS $\hat{\imath}\hat{\tau}_{8,3}\hat{X}_{8,-1}$, that is it upholds the equation:

$$-\vec{E}\left(t + 3\frac{2\pi}{8\omega}, X - \frac{2\pi}{8k_x}, Y\right) = \vec{E}(t,X,Y) \tag{28}$$

The microscopic inversion operation flips the handedness of chiral field. Therefore, following similar argument as in the previous case, the induced polarization of the two enantiomers will have the relation:

$$\hat{\imath}\hat{\tau}_{8,3}\hat{X}_{8,-1}\vec{P}^L(t,X) = \vec{P}^R(t,X) \tag{29}$$

Which according to the multi-scale DS theory, specifically Eq. (7) in the main text, leads to:

$$-\vec{F}^L(q_1\omega, q_2k_x)e^{\frac{2\pi i(3q_1 - q_2)}{8}} = \vec{F}^R(q_1\omega, q_2k_x) \tag{30}$$

Inserting the relation to CD:

$$CD(q_1\omega, q_2 k_x) = 2\frac{\left|\vec{F}^L(q_1\omega, q_2 k_x)\right|^2 - \left|\vec{F}^R(q_1\omega, q_2 k_x)\right|^2}{\left|\vec{F}^L(q_1\omega, q_2 k_x)\right|^2 + \left|\vec{F}^R(q_1\omega, q_2 k_x)\right|^2}$$

$$= 2\frac{\left|\vec{F}^L(q_1\omega, q_2 k_x)\right|^2 - \left|-\vec{F}^L(q_1\omega, q_2 k_x)e^{\frac{2\pi i(3q_1-q_2)}{8}}\right|^2}{\left|\vec{F}^L(q_1\omega, q_2 k_x)\right|^2 + \left|-\vec{F}^L(q_1\omega, q_2 k_x)e^{\frac{2\pi i(3q_1-q_2)}{8}}\right|^2} \quad (31)$$

$$= 2\frac{\left|\vec{F}^L(q_1\omega, q_2 k_x)\right|^2 - \left|\vec{F}^L(q_1\omega, q_2 k_x)\right|^2}{\left|\vec{F}^L(q_1\omega, q_2 k_x)\right|^2 + \left|\vec{F}^L(q_1\omega, q_2 k_x)\right|^2} = 0$$

Leads to forbidden CD.

Thus, our theory for multi-scale DSs can also be applied in the field of chiral light-matter interactions to explain ultrafast chiral dichroism in solid and molecular systems.